\documentclass[final]{svjour2}
\usepackage{rotating}
\usepackage{amssymb}
\usepackage{mathptmx}
\usepackage{gensymb}
\usepackage[numbers]{natbib}
\graphicspath{{Figure/}}

\makeatletter
\journalname{Journal of Low Temperature Physics}

\bibpunct{[}{]}{,}{n}{}{,}

\begin{document}

\newcommand{\hdblarrow}{H\makebox[0.9ex][l]{$\downdownarrows$}-}
\title{Grand Challenges in Quantum Fluids and Solids}

\author{Y. Lee$^1$ \and W.P. Halperin$^2$}

\institute{$^1$ Department of Physics, University of Florida, Gainesville, FL 32611, USA\\
\email{yoonslee@phys.ufl.edu}\\
$^2$ Department of Physics and Astronomy, Northwestern University, Evanston, IL 60208}

\maketitle

\begin{abstract}
We present a short review of the current and recent development in the field of superfluid $^{3}$He and $^{4}$He. The review is based on the topics discussed at the NSF sponsored Workshop on Grand Challenges in Quantum Fluids and Solids chaired by Moses Chan from Pennsylvania State University and held at University of Buffalo in August, 2015. 

\end{abstract}

\section{Introduction}
In pursuing sources for newly discovered argon, William Ramsey conducted a spectrospcopic test in 1895 on the gas extracted from cleveite purchased from a local source. Instead of argon, what he found was an unknown element that he called crypton -- until it was eventually identified as helium, a known element in the solar atmosphere \cite{Seibel}.  This light gas attracted the interest of many scientists including Kammerlingh Onnes in Leiden.  He used helium in his quest for absolute zero and finally succeeded in liquefying it on July 10th of 1908, and soon thereafter he reached temperatures below 1~K.  This monumental event directly led to the discoveries of superconductivity in mercury and superfluidity in helium, opening the two major fields in modern low temperature physics –- superconductivity and quantum fluids and solids.  After WW~II small quantities of helium three became available and rapidly the low temperature properties of helium turned into the subject of intense research, providing a basis for understanding interacting bosonic ($^4$He) and fermionic ($^3$He) many-body systems in the form of extremely pure liquid.

In parallel, researchers have continued to pursue ever lower temperatures developing new cryogenic techniques (and inevitably thermometry) such as Pomeranchuk cooling, the dilution refrigeration, and adiabatic demagnetization of nuclear spins \cite{Pobell}.  Over the past century, the lowest attainable temperature (energy) has been reduced from $\sim 1$~K ($10^{-4}$~eV) to $\sim 0.1$~nK ($10^{-14}$~eV), a remarkable 10 orders of magnitude, which mirrors the progress in achieving higher energies in particle physics during the same time span (from $10^{6}$~eV to $10^{16}$~eV) (Figure 1).  In the past five decades the technical advancement of dilution refrigeration has made millikelvin temperatures widely accessible.  Nowadays, with only mild exaggeration, one can reach and stay below 10~mK with a single touch of the screen without going through the ritual of transferring cryogens. The sustainable and low-maintenance millikelvin systems have blossomed into a billion dollar industry and are well positioned as the foundation for research in diverse areas in science and engineering extending beyond the traditional low temperature physics, including dark matter searches and quantum information processing, the latter with great commercial potential.

\begin{figure}[t]
  \includegraphics[height=3in]{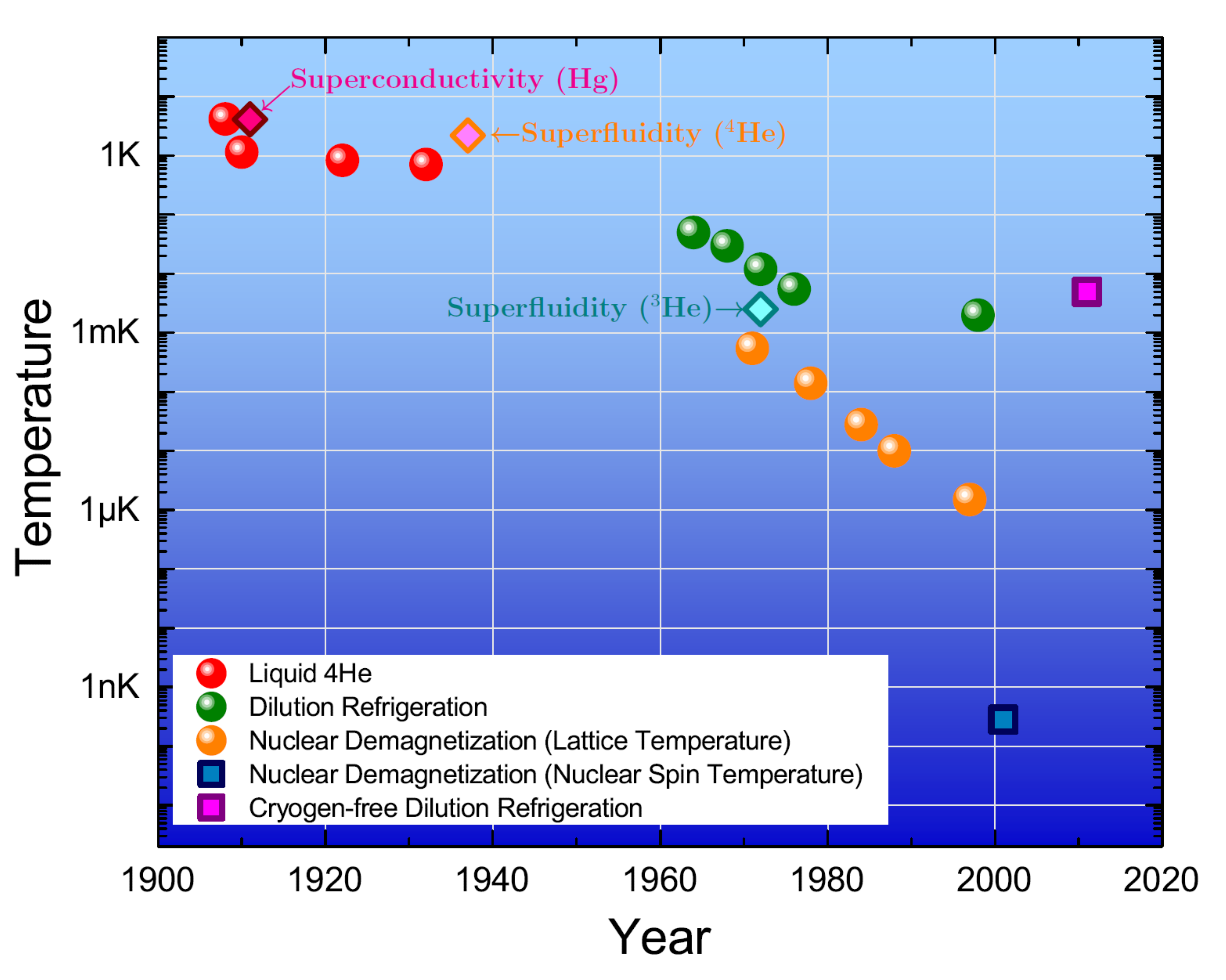}
\caption{The lowest temperature reached since the liquefaction of helium in 1908 by Kammerlingh Onnes using various cryogenic techniques along with the discoveries of superconductivity and superfluidity (diamonds).  Cryogen-free dilution refrigerators reaching below 10~mK are now commercially available(red square).}
\end{figure}

The research in quantum fluids and solids covers diverse topics  beyond superfluidity itself including the fascinating quantum properties of the solid-liquid interface of $^4$He and quantum crystal growth, the magnetic properties of solid $^3$He stemming from quantum tunneling of $^3$He atoms, and quantum turbulence.   
In particular, solid $^4$He has been in the limelight in recent years.  Kim and Chan's experiment in 2004 was interpreted as evidence for a supersolid, generating an enormous amount of excitement and activity \cite{Kim}.  Although the original claims have been under intense scrutiny, there are still lingering questions that remain unresolved.  For interested readers we refer readers to the special issue of this journal (Vol.~180, 2015) and the recent article by Hallock \cite{Hallock}.  Here, we focus on the superfluid phases of $^4$He and $^3$He and provide an overview of current status and recent progress with a forward-looking perspective. The topics discussed in this article are selected based on the discussions during the workshop on the frontiers and opportunities in quantum fluids and solids held at the University of Buffalo in summer of 2015. 

\section{Basics of Superfluid Helium}
Superfluidity of the two isotopes sets in at drastically different temperatures: around 2~K in $^4$He and 2~mK in $^3$He.  The difference of one tiny neutron between them is responsible for their very different quantum behavior.  While the superfluidity in $^4$He is closely related to the Bose-Einstein condensation, the superfluid $^3$He is a result of forming bosonic bound pairs of two fermions -- Cooper pairs -- as in superconductors.  The two-fluid model introduced by Tizsa and Landau provides a phenomenological framework in understanding many enchanting properties of superfluids in general. Below the superfluid transition temperature, one can view the liquid as having two interpenetrating fluids: the inviscid superfluid and the viscous normal fluid, each moving with their own velocities $\vec{v}_{s}$ and $\vec{v}_{n}$, respectively.  The normal fluid component represented by its density $\rho_{n}$ decreases as the temperature falls and eventually disappears at zero temperature leaving only the superfluid component $\rho_{s}$.  The normal component is basically the gas of elementary excitations thermally populated from the ground state and therefore carries the entropy and viscosity.   These are phonons and rotons in $^4$He and fermionic quasiparticles in $^3$He.  It may surprise the readers to note that right above the superfluid transition liquid $^3$He is as thick as olive oil while $^4$He has lower viscosity than water. 

Simple superfluids -- $^4$He and conventional superconductors belong in this category -- are described by a macroscopic quantum wavefunction $\Psi = |\Psi|e^{i\phi}$.  The superfluid velocity is then related to the gradient of the phase, $\vec{v}_{s} = \frac{\hbar}{M} \nabla\phi$, which makes the superflow irrotational, $\nabla \times \vec{v}_{s} = 0$.   Here $M$ represents the mass of the relevant boson: either the mass of a $^4$He atom or two $^3$He atoms. One of the immediate consequences of a quantum irrotational fluid is the quantized vortex, a topological defect surrounded by an azimuthal superflow field. The circulation of the superflow is quantized in units of $\kappa = \frac{h}{M}$, reminiscent of the magnetic quantum flux $\Phi_{o} = \frac{h}{2e}$ in superconductors. 

The wavefunction of a Cooper pair in $^3$He is a little more complicated.  Helium three atoms form a bound pair with the total spin and orbital angular momenta both equal to $\hbar$ ({\it spin-triplet, p-wave}) rather than zero as in the conventional superconductors, which makes the Cooper pair magnetic and anisotropic with a non-trivial internal structure \cite{Leggett}.  In fact superfluid $^3$He is the first known unconventional pairing condensate. Being an isotropic non-magnetic fluid, the normal state has the maximal symmetry but in the transition to superfluidity various symmetries are spontaneously broken in addition to gauge symmetry, giving rise to impressively diverse physical phenomena. Since then, many new superconducting materials have been discovered which also break symmetries of the normal state and are in this general class of unconventional pairing symmetry.

The Cooper pair wavefunction of superfluid $^{3}$He is represented by a linear superposition of 9 components of a {\it p-wave, spin-triplet} wavefunction: $\Psi = \sum_{L,S}C_{L,S}\psi_{L,S}$ where $C_{L,S}$ is a complex number with $L,S = -1,0,+1$. The most convenient way of representing this structure is through the order parameter in the form of a $3\times 3$ complex matrix.  The element of this matrix, $A_{\mu j}$, consists of two indices representing the spin ($\mu$) and the orbital ($j$) spaces. The high number of degrees of freedom in superfluid $^{3}$He allows for the possibility of multiple superfluid states which have distinct gap structures as shown in Figure 2.  The difference in the radial distances between the blue inner sphere -- Fermi surface in momentum space -- and the red outer surface indicates the size of the gap in that orientation.  The A-phase (left) has two point nodes at the opposite poles of the Fermi surface while the B-phase (middle) has an isotropic gap.  The polar phase (right) has a line node on the equator and was recently discovered for $^{3}$He in anisotropic aerogel.  

While the above gap structure manifests the anisotropic nature of the superfluid states, the detailed information of the symmetry of each phase is reflected in the order parameter:  
$$
	A_{\mu j}^{A} = \Delta_{A}\hat{d}_{\mu} \left(\hat{m}_{j} + i\hat{n}_{j}\right)e^{i\phi},
$$
$$
	A_{\mu j}^{B} = \Delta_{B}R_{\mu j}e^{i\phi},
$$
$$
	A_{\mu j}^{P} = \Delta_{P}\hat{d}_{\mu}\hat{n}_{j}e^{i\phi}. 
$$

\noindent These order parameters reveal their anisotropic nature in the form of vectors in both spin ($\hat{d}$) and orbital ($\hat{m}$, $\hat{n}$, and $\hat{\ell}$) spaces.  In general, these vector fields can have spatial variations called textures.  The orbital vectors for the A-phase determine the direction of the angular momentum through $\hat{\ell} = \hat{m} \times \hat{n}$: $\hat{m}$, $\hat{n}$, and $\hat{\ell}$ are mutually orthogonal. The appearance of the relative phase between $\hat{m}$ and $\hat{n}$ implies that the overall phase of the Cooper pair is jointly influenced by the change in $\phi$ and the rotation around $\hat{\ell}$.  On the other hand, $R_{\mu j}$ in the B-phase, a relative spin-orbit rotation matrix, implies locking between the two spaces.  The order parameter structure allows many different classes of topological structures such as vortex lines and dysgyrations \cite{Volovik1976}. 

\begin{figure}[h]
  \includegraphics[height=3in]{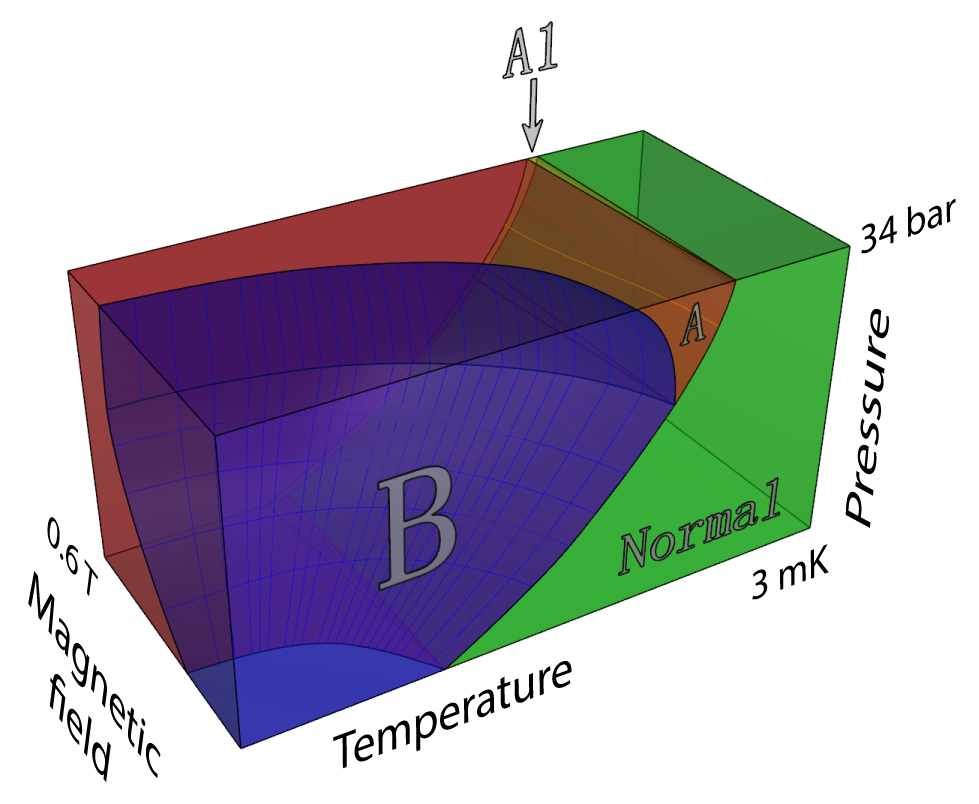}

  \includegraphics[height=1in]{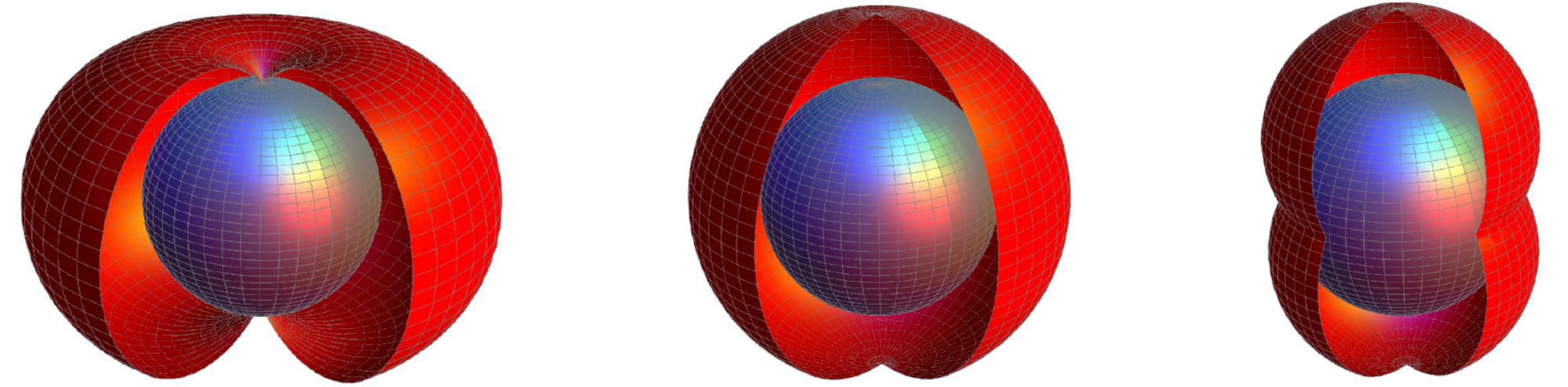}

\caption{Temperature-pressure-magnetic field phase diagram of superfluid $^{3}$He and the gap structure. The difference in the radial distances between the blue inner sphere -- Fermi surface in the momentum space -- and the red outer surface indicates the size of the gap in that orientation.  The A-phase (left) has two point nodes at the opposite poles of the Fermi surface while the B-phase (middle) has an isotropic gap.  The polar phase (right) has a line node on the equator.}
\label{fig:test}
\end{figure}

\section{Superfluid Helium 4}
\subsection{Quantum Turbulence}
Turbulence in an ordinary fluid arises in the regime of a high Reynolds number -- the ratio of the inertial force to the viscous force -- where the presence of the viscous force is essential.  It was an intriguing question whether turbulence could also exist in an inviscid superfluid.  In 1955, Feynman proposed the possibility of superfluid turbulence or quantum turbulence arising from a tangle of quantized vortex lines as an explanation for the experimental observation of dissipative thermal counterflow -- counterflow of two fluid components in the presence of a heat flux -- at a high relative velocity $|\vec{v}_{n} - \vec{v}_{s}|$ \cite{Feynman1955}.  The experimental confirmation followed immediately and numerous experimental and theoretical studies revealed interesting features unique to the quantum turbulence (QT) generated by thermal counterflow \cite{Barenghi}.  It was natural to think that QT would be quite different from classical turbulence (CT) considering the unique way of generating QT in the form of stable identical quantized vortices rather than continuous eddies as in CT, not to mention the two interpenetrating fluids that can interact via the nonconventional mutual friction force \cite{Skrbeck}.

However, the QT experiments conducted in the mid-90's revealed surprising similarity to CT and have shifted the main interest in this field to understanding quantum turbulence in relation to its classical counterpart.  When the turbulence in the superfluid was generated by the methods conventionally used in CT such as counter rotating cylinders or moving grids, researchers found a Kolmogorov energy spectrum, a robust statistical law governing CT: $E(k) \sim \epsilon^{2/3}k^{-5/3}$ where $\epsilon$ is the energy flux and $k$ is the wavenumber.   Kolmogorov scaling tells how the kinetic energy in turbulence is distributed at different length scales.   Above 1~K where a substantial amount of normal fluid is present, it is believed that the two fluids are coupled by the mutual friction at scales greater than the mean inter-vortex distance, $\ell_{v}$. In this case, the fluid behaves like a single-component viscous fluid, exhibiting quasi-classical features. At small scales below $\ell_{v}$, the two fluids are no longer coupled. Dissipation then sets in due to both the viscosity of the normal fluid and mutual friction \cite{Barenghi}. The similarity between QT and CT suggests the possibility of using the superfluid to gain deeper understanding of turbulence in general, and additionally its low viscosity provides the opportunity for generating flows with extremely large Reynolds number ($\gtrsim 10^{8}$) on a rather moderate experimental scale. 

\begin{figure}
  \includegraphics[height=4in]{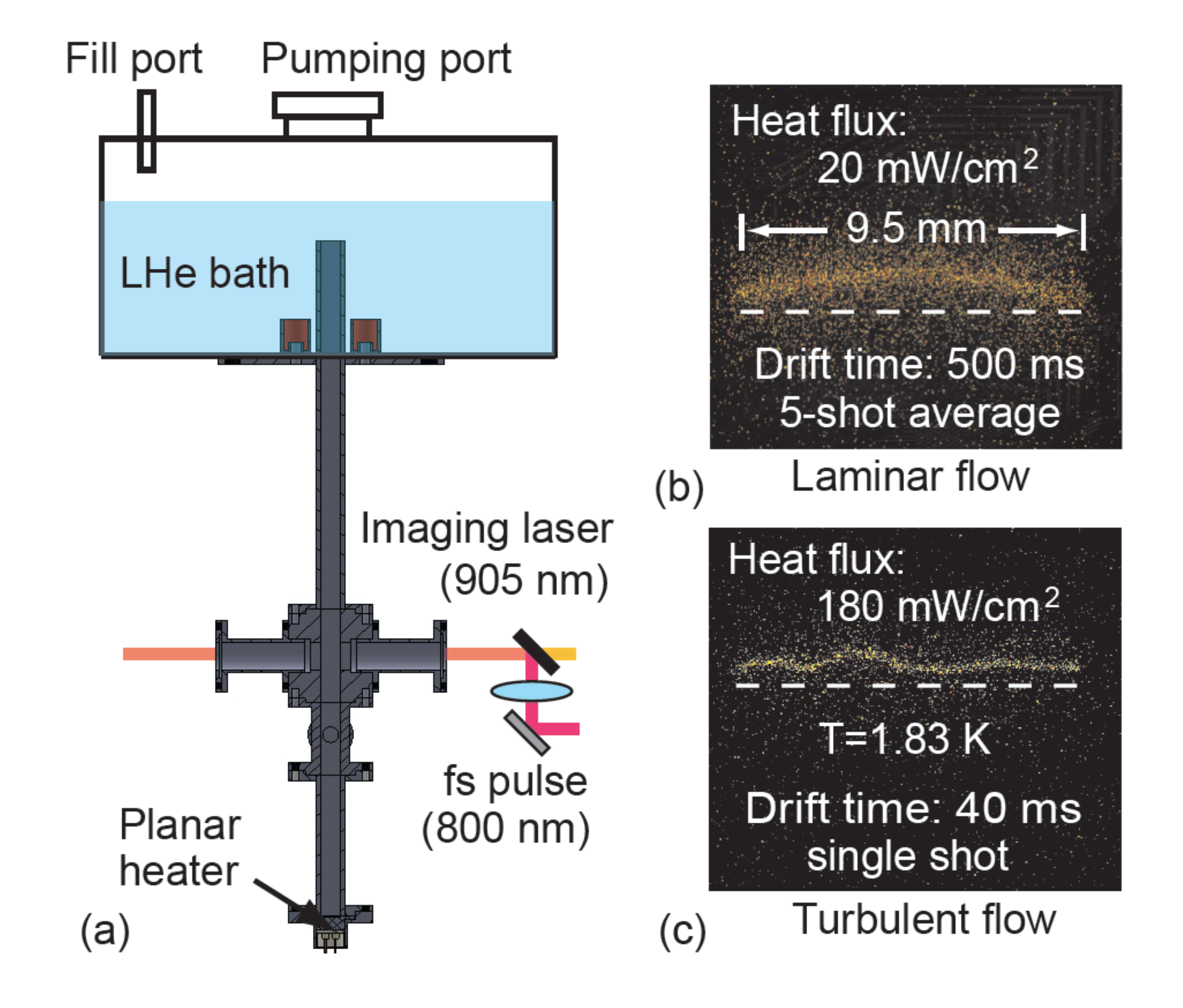}
  
\caption{(a) Schematic diagram of the experimental setup for flow visualization using He$_{2}^{*}$ molecules. The vertical square channel is connected to the top bath filled with superfluid helium. A high intensity femto-second laser (red beam) through the windows ionizes helium and creates a tracer line of He$_{2}^{*}$ molecules. Then the imaging laser at 905~nm (yellow beam) drives the tracers to produce fluorescent light (640~nm) for the imaging. (b) An image of the tracer line in thermal counterflow at a low heat flux to the top.  The initial tracer line, indicated by the dashed line, deforms into a nearly parabolic shape, indicating a laminar Poiseuille velocity profile of the normal component. (c) A single-shot image of the tracer line at a substantially larger heat flux. The tracer line deforms randomly, indicating turbulent flow in the normal fluid. Through the analysis of multiple sequential images, important statistical quantities such as velocity probability function can be obtained. (Courtesy of Wei Guo at Florida State University) }
\end{figure}

From the experimental point of view, it is essential to have quantitative measurements of the velocity fields with adequate temporal and spatial resolutions. Exciting progress was made at Yale, University of Florida, and Florida State University, developing powerful visualization techniques using long lifetime excited He$_{2}$* triplet molecules as tracers imaged via laser-induced fluorescence (Figure 3) \cite{Guo, Marakov}. This followed earlier pioneering work by Zhang and Van Sciver \cite{Zhang} who used micron-sized polymer particles as the tracers to visualize the large-scale counterflow turbulence, and the work by Lathrop's group at the University of Maryland who successfully imaged the vortex lines and their reconnection using frozen hydrogen particles of similar sizes \cite{Bewley}.  

Quantum turbulence has also been investigated in superfluid $^3$He \cite{Fisher2001, Finne}. In 2001 Fisher {\it et al.} at Lancaster University were able to generate and detect the turbulence in the B-phase of $^3$He using vibrating wire resonators, which have been a major tool, allowing confirmation of the Kolmogorov energy spectrum at long length scales.  It has been speculated that vortices should play an important role in $^3$He because of the large core size ($\sim 10$~nm for $^3$He and 0.1~nm for $^4$He) and the localized excitations in the vortex core \cite{Silaev}.  A central question to be resolved is to find the mechanism for energy dissipation at small length scales.  Researchers at Lancaster University have been working on a visualization technique based on a unique process in a fermionic superfluid, Andreev reflection: a retro-reflection involving the branch conversion between the particle and hole \cite{Fisher}.  A quasiparticle can be reflected by a vortex line due to the potential barrier induced by the superflow around the vortex, giving rise to a large cross-section and consequently its shadow in the quasiparticle flux.   Currently, a few groups at Lancaster University \cite{Baggaley}, Insitut Louis Neel, and University of Florida \cite{Gonzalez} are developing techniques to image the shadows from the vortex lines using various types of mechanical oscillators in the form of arrays of miniature quartz tuning forks and micro/nano-electro-mechanical oscillators serving as the pixels in a quasiparticle camera.

\subsection{Ultra-high Sensitive Superfluid Detection}
Condensed matter systems at low temperatures offer desirable platforms for sensitive, high energy-resolution particle detection including neutrinos or dark matter. Concrete schemes for cryogenic detection bloomed in the 80's and some of them are now at work in the millikelvin range, CDMS, EDELWEISS, CRESST, and CUORE, to name a few. Aside from the unmatched energy sensitivity that superfluid phases of helium offer, their extreme purity -- surpassing the semiconductor materials employed in the current detectors -- have made them attractive media for this type of application. On the other hand  the  macroscopic coherent state of superfluidity is a perfect fit for interferometry as has already become a standard technique in the superconducting quantum interference device (SQUID).  The superfluid version, the SHeQUID, has been built and tested. While a SQUID is the most sensitive magnetic flux detector, a SHeQUID can be configured as an extremely sensitive rotation detector. The detector built by Sato and Packard at Harvard demonstrated its potential to achieve very high rotational sensitivity $\approx  10^{-8}~rad/s/\sqrt{Hz}$ \cite{Sato}. 

In superfluid $^4$He the longitudinal sound mode has extremely low damping at low temperatures, which has been projected to have an acoustic quality factor of $Q \approx 5 \times 10^{11}$ at 10~mK.  De Lorenzo and Schwab at Caltech \cite{Schwab} integrated this high $Q$ mechanical system into a high $Q$ microwave cavity to form an opto-mechanical device.  The modulation of fluid density from the acoustic motion should result in the modulation of permittivity, shifting the microwave cavity resonance.  At around 30~mK, they demonstrated the detection of acoustic modes with a quality factor as high as $\approx 10^{7}$ in a device composed of a high quality superconducting microwave cavity filled with superfluid. If configured as an inertial detector, it could reach a strain sensitivity of $s \approx 10^{-26}$ under the assumption of $Q \approx 10^{11}$ at 10~mK and thermal noise limited detection \cite{Schwab}.  The detection of the gravitational wave from the binary black holes in 2016 was at $s \approx 10^{-21}$ using two 4~km double-armed interferometers at Advanced LIGO.  One might imagine superfluid gravitational wave detectors of a Rubik's cube size in multiple laboratories.

\subsection{Qubits on Helium}
The surface of superfluid helium in the millikelvin range is indeed impurity free, atomically smooth on a macroscopic scale, and interfaced to what is practically vacuum.  It is an atomically perfect flatland with no obstacles to interfere with the motion of an electron placed on the surface where it feels a weak attractive force. In fact, the combination of the image charge and the high potential barrier of the surface presents a 1D hydrogen-like potential along the surface normal direction. Therefore, electrons form bound states while moving freely within the plane, an almost perfect 2D electron gas. At low temperatures the only significant scattering of an electron comes from the thermally excited surface fluctuations called ripplons. This coupling is so weak that the electrons on helium have the highest mobility of any physical system exceeding 10$^8$~cm$^2$/Vs.  So, electrons on helium is an exceptional platform for studying 2D many body phenomena. Although one cannot reach the quantum degenerate regime due to an instability at high density, a classical 2D Wigner solid -- an electron crystal -- has been observed and investigated extensively \cite{Dahm}.   

In recent years, the major effort in this area have been steered toward a rather multi-disciplinary topic, quantum information processing, following the compelling proposition by Platzman and Dykman to use the two lowest energy states of electrons on helium as a qubit \cite{Platzman}.  The weak coupling to the environment results in a very long coherence --  as long as $\approx 100$~s when the spin state of an electron is used.  Furthermore nano-fabrication technology can be adopted to build scalable qubits in this system.  Regardless of the scheme, it is crucial to have means to control and manipulate individual electrons. Lea's group at University of London was the first to demonstrate this capability in 2001 \cite{Glasson}. They created a Wigner wire on helium, capillary-condensed into a micron-sized channel.  Using fish-bone type gate electrodes, they were able to manipulate the solid electrons. Since then, variations of the technique have been developed by several groups including RIKEN \cite{Shirahama}, Princeton, and the University of Chicago \cite{Yang}.  Recently, Lyon and his collaborators at Princeton \cite{Bradbury} used charge-coupled device (CCD) gates to successfully transport packets of electrons, to the level of a single electron per pixel, simultaneously along the 120 micro-channels with an exceptional efficiency of 0.999,999,999,9. 

\section{Superfluid Helium 3}

\subsection{Symmetry and Multiple Superfluid Phase}
There are two superlfuid phases in zero magnetic field, the A- and the B-phase with a spin superfluid state called the A$_{1}$-phase that opens up with a high magnetic field.  In addition to broken gauge symmetry, the A-phase breaks spin and orbital rotation symmetry, and has chiral symmetry.  There is a remarkable similarity between the pressure-temperature phase diagram of $^{3}$He and the magnetic field-temperature phase diagram of the unconventional heavy fermion superconductor UPt$_{3}$, each having multiple superfluid phases where one of them is of chiral symmetry. The common aspects of unconventional superconductivity and superfluid $^{3}$He are currently prime topics of research \cite{Norman}. While the number of experimental results supporting the chiral nature in other superconductors is growing,  Kono and colleagues at RIKEN in Japan garnered indisputable experimental evidence of chiral symmetry in the A-phase using electron bubble transport just below and parallel to the free superfluid surface \cite{Ikegami}.  In a perpendicular magnetic field this ion transport is subject to a transverse Magnus force from the orbital supercurrent around the bubble, which depends on the sign of the superfluid orbital angular momentum normal to the surface. In this way macroscopic regions of the superfluid were identified as having a single chiral domain. 

In contrast, the characteristic symmetry of the B-phase order parameter is an unusual broken relative spin-orbit rotation symmetry.  A direct demonstration of this symmetry was achieved by the authors, and their collaborators at Northwestern \cite{Lee}, through experimental realization of the acoustic Faraday effect.  Transverse sound, which in general does not exist in fluids, propagates in the B-phase.   When a magnetic field is applied along the propagation direction, the linear polarization of the sound wave rotates in the same manner as in magneto-optical Faraday rotation.  This effect, which was theoretically predicted by Moores and Sauls at Northwestern \cite{Moores}, is the direct manifestation of spin-orbit locking due to the broken symmetry in this phase. This transverse technique has been used as a tool to better understand the spectroscopy of order parameter collective modes and Andreev surface bound states.

In addition to the fermionic excitations in $^3$He mentioned above, there are bosonic excitations associated with 18 components of the order parameter called the order parameter collective modes \cite{Halperin}.  The spectra of these excitations directly reflect the symmetry of the wavefunction and have played a crucial role in determining its structure (symmetry). Some of the branches do not have an excitation gap -- a massless Dirac particle or a Nambu-Goldstone mode emerging from broken continuous symmetry -- like acoustic phonons in solids arising from broken translational symmetry.  However, many of them have an excitation gap, and the dispersion for these branches is exactly that of a  massive Dirac particle, $\omega^{2}(k) = \omega_{o}^{2} + c^{2}k^{2}$.  This aspect has been discussed in terms of the Higgs mechanism in the Standard Model \cite{Volovik13, Sauls17}. A recent experiment by a group of researchers at Aalto University and the Landau Institute beautifully demonstrated how a gapless mode (a massless particle) in the B-phase becomes gapped (gains mass) on the energy scale much smaller than the condensation energy, an interesting analogy to the Higgs boson discovered in 2012 at CERN at a much lower energy (125~GeV) than the expected energy scale of 1~TeV \cite{Higgs}.  

\subsection{Engineering New Phases of Superlfuid Helium 3}
\noindent The influence of boundaries or disorder has particular significance for any unconventional superconductors since all forms of quasiparticle scattering can break pairs or modify the superconducting state.  The discovery at Cornell \cite{Porto} and Northwestern \cite{Sprague} that sufficiently high porosity silica aerogel can host uniform phases of superfluid $^3$He opened a path to study systematically the role of static disorder in such system \cite{Aerogel}. The $^{3}$He system has the advantages of continuous tuning of the superfluid coherence length, from 15~nm to 80~nm through changes of pressure, while holding the impurity framework constant.  Measurements of acoustics, magnetization, NMR frequency shifts, thermal conductivity, and heat capacity concur that the two superfluid phases, A and B familiar from bulk $^3$He are suppressed in arogel as expected by theories based on homogeneous isotropic scattering \cite{Thuneberg}.  However, the experimental and theoretical studies in the past 10 years revealed that anisotropic disorder has significant influence on the structure of the superfluid phase because of the anisotropic nature of the superfluid itself.  Silica aerogels that are homogeneous and isotropic on the scale of the quasiparticle mean-free-path, for example, 150 nm for a 98\% porosity aerogel have a dramatic effect destabilizing the A-phase in favor of the B-phase which then fills the entire phase diagram at zero magnetic field \cite{Vicente}.  On the other hand anisotropic aerogel samples, \emph{stretched} during growth by promoting radial shrinkage,  were found to stabilize the chiral A-phase at all pressures. 2017 Fritz London Memorial Prize recognized the contributions of Halperin, Parpia, and Sauls in deepening our understanding of the effects of disorder in superfluid $^{3}$He.

At the Kapitza Institute in Moscow, Dmitriev and colleagues have found another type of aerogel which has much greater anisotropy than can be obtained with silica aerogel \cite{Dmitriev}.  The material, called Obninsk or Nafen shown in Figure 3, consists of nematically-ordered Al$_{2}$O$_{3}$ strands. Their detailed NMR \cite{Dmitriev} and torsional oscillator \cite{Zhelev} studies showed the superfluid phase in nematic aerogel is not the A nor the B but a completely new phase, the polar phase, for which the energy gap structure is shown in Fig.~1. A collaboration of the groups at Aalto University and the Kapitza Institute have performed NMR measurements on $^3$He in a Nafen sample containing the new polar superfluid phase during physical rotation at sub-millikelvin temperatures \cite{Autti}.  Under these conditions they formed the elusive half quantum vortex (HQV) with $\frac{1}{2}\kappa$ circulation that is predicted to harbor unpaired Majorana excitations at its core (Figure 3). There is also evidence for this vortex in several unconventional superconductors and Bose-Einstein condensates. The prediction of this state by Volovik and Mineev \cite{Volovik1976} led to their award of the Lars Onsager Prize of the American Physical Society in 2014.
\begin{figure}
  \includegraphics[height=3.5in]{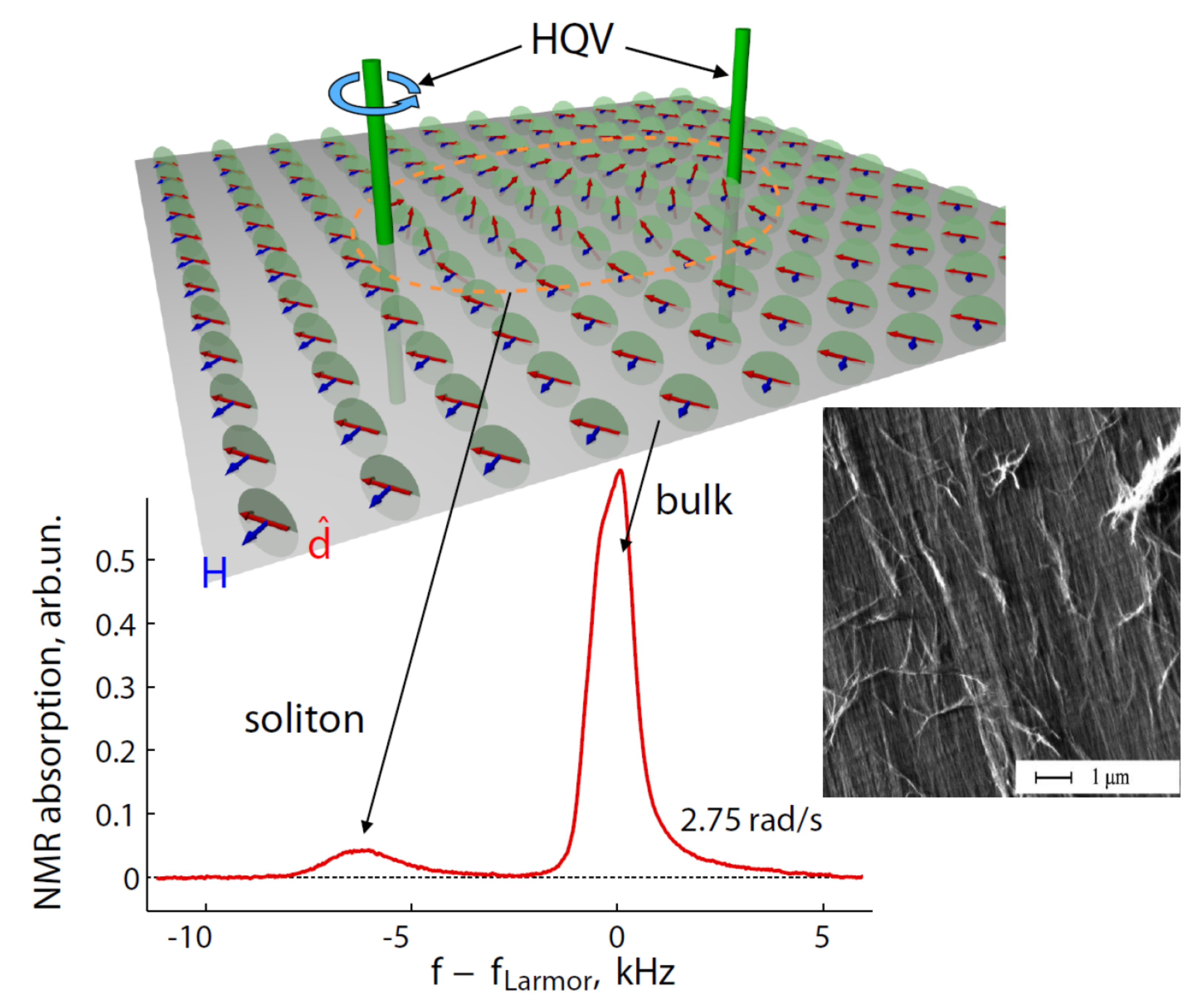}
  
\caption{Top: A pair of HQVs connected by a soliton $\hat{d}$ (red arrow) texture in the polar phase under rotation. Both the rotation axis and the Nafen strands are aligned in the $\hat{z}$-direction. In a magnetic field (blue arrow), $\hat{d}$ (red arrow) is confined in the plane whose surface normal is in the direction of the magnetic field  (blue arrow), and $\hat{d}$ rotates by $\pi$ when circling around the HQV, which compliments the $\kappa/2$ circulation to guarantee the single valuedness of the wavefunction. The above vortex structure is formed when a transverse magnetic field is applied after cooling into the polar phase under rotation in zero magnetic field. Bottom: The negative NMR frequency shift of the satellite peak with respect to the main bulk peak demonstrates the existence of spin wave modes bound to the soliton emerging between the neighboring HQVs. (Adopted from Ref.~\cite{Autti}) }
\label{fig:test}
\end{figure}

These findings lay out an intriguing way of engineering novel superfluid states and  excitations using anisotropic structures or confinement.   Further theoretical and experimental work on $^3$He in confinement in slabs, cylindrical pores, and aerogels is being pursued in a number of laboratories today to extend our knowledge of how correlated structures or confinement affect unconventional pairing.  Fabrication of micron and sub-micron scale structures for $^3$He promises dramatic effects on the thermodynamics of the superfluid, stabilizing new phases with different order parameter symmetry.  For example Vorontsov and Sauls \cite{Vorontsov} predicted that $^{3}$He confined in a thin slab would stabilize a phase of the superfluid with periodic in-plane modulation of its order parameter, analogous to the Fulde-Ferrell-Larkin-Ovchinnikov state, recently reported for example in an organic superconductor \cite{Mayaffre}.  The theoretical work indicates that confinement can lead to spontaneously broken translational symmetry in a superfluid, a remarkable phenomenon being sought in superfluid $^{3}$He by the collaboration between the research groups of Parpia at Cornell and Saunders at Royal Holloway \cite{Levitin}.   

\subsection{Topological Phenomena}
 A strong continuing interest in topologically non-trivial materials stems in part from the goal of creating qubits.  One class of such systems are the Majorana states which have been demonstrated in nanowires and sought after at the surface of bulk topological insulators.  It is well established theoretically that the surface of the B-phase and vortex cores in the A-phase and the recently discovered polar phase are host to these Majorana excitations. Downstream application to quantum information processing might not be realizable in $^3$He. Nonetheless, $^3$He is a valuable platform for coupling theory and experiment to better understand basic principles of these topologically protected quantum states in systems where the order parameter structure is well-established and the experimental tools are refined and precise.  Currently, there are international efforts to develop experimental techniques to provide further evidence for these Andreev bound states.  These methods include measurements by the group at the Tokyo Institute of Technology \cite{Aoki} of the acoustic impedance of transverse sound at the interface between a quartz transducer and the superfluid B-phase, exactly where these states are localized.  Heat capacity measurements of the bound states have been reported \cite{Choi06}, perhaps uniquely possible in the quantum fluid paradigm, and are continuing in several laboratories.  A novel and ultra-clean superfluid surface lies at the equilibrium interface between A and B-phases which can be stabilized and manipulated. Additionally, a new class of instrumentation has been developed to investigate the surface physics of superfluid $^3$He, using micro/nano-electro-mechanical devices at the University of Florida \cite{Pan}.  It is clear that there are rich opportunities for exploring these localized quantum states in $^3$He by different techniques, and that they have immediate connection to related phenomena in other condensed matter systems.

\section{Conclusion}
There is no doubt that the turn-key dilution system has widened the scope of research beyond traditional low temperature physics and has attracted researchers from outside of the quantum fluids and solids community to take advantage of the exceptional physical properties of these quantum systems for application to other science and new technology. There is even a thrust with commercial support to extend present day cryogenic platforms to reach sub-millikelvin temperatures.  With the ability to cool  electronic systems to the microkelvin range and potential of using superfluid $^3$He as a tool for other scientific research, low temperature physics has a rich and challenging future.

\begin{acknowledgements}
This work is supported by NSF through DMR-1205891 (YL) and DMR-1602542 (WPH).
\end{acknowledgements}


\end{document}